\documentclass[prd,12pt,nofootinbib,preprint,floatfix]{revtex4}

\usepackage{amsmath,amssymb}
\usepackage{graphicx}

\usepackage{setspace}
\usepackage{color}

\begin{document}

\title{Free-form smearing for bottomonium and $B$ meson spectroscopy}

\author{Mark Wurtz$^1$, Randy Lewis$^1$ and R. M. Woloshyn$^2$}
\affiliation{$^1$Department of Physics and Astronomy, York University, Toronto,
             Ontario M3J 1P3, Canada\\
             $^2$TRIUMF, 4004 Wesbrook Mall, Vancouver, British Columbia V6T 2A3, Canada}

\begin{abstract}
To obtain high-quality results from lattice QCD, it is important to use operators that produce good signals for the quantities of interest.  Free-form smearing is a powerful tool that helps to accomplish that goal.  The present work introduces a new implementation of free-form smearing that maintains its usefulness and reduces its computational time dramatically.  Applications to the mass spectrum of $B$, $B_s$, $B_c$ and bottomonium mesons show the effectiveness of the method.  Results are compared with other lattice QCD studies and with experimental data where available. The present work includes the first lattice QCD exploration for some of these mesons.
\end{abstract}

\maketitle

\section{Introduction}

In the usual Lagrangian formulation of lattice quantum chromodynamics (QCD) physical quantities are extracted from Euclidean-time correlation functions of operators which serve as interpolating fields for the states of interest. In general, many states contribute to a given correlation function. Contemporary applications of lattice QCD carried out to study radial and orbital excitations of hadrons can benefit from methods that can enhance signals in a correlation function from particular states selectively. This is what free-form smearing aims to do. 

A common approach to operator building is to begin with a gauge-invariant operator defined at a single lattice site (or a small number of sites) that has the desired $J^{PC}$ quantum numbers, and then to smear that operator spatially through an iterative Gaussian function \cite{Gusken:1989ad,Alexandrou:1990dq,Allton:1993wc}.  At the time this method was developed the goal was to suppress excited-state contributions to the correlation functions. As pointed out in Ref.~\cite{vonHippel:2013yfa}, this approach can require many iterations, especially on fine lattices, and the resulting operator is typically not Gaussian shaped.  This led the authors of Ref.~\cite{vonHippel:2013yfa} to propose a method called free-form smearing, which can produce a gauge-invariant operator of any desired shape. The new method still uses an iterative Gaussian approach but combines it with a reweighting formula involving a function that the user can choose arbitrarily.
This allows one to enhance or reduce the contribution of particular
states to the correlation function.
Analyzing simultaneously correlators constructed from operators with different free-form smearing profiles allows for a more reliable separation of contributions of different states. At present, free-form smearing is practical at the source and not at the sink, but it still provides a clear improvement over traditional Gaussian smearing, as will be demonstrated in the present work.

Our implementation of free-form smearing differs from that of Ref.~\cite{vonHippel:2013yfa} by avoiding the iterative Gaussian step entirely.  On an $L^3\times T$ lattice, this provides a speedup of order $L$.  This version of free-form smearing is introduced in Sec.~\ref{sec:FFsmearing}.

For a demonstration of the power of free-form smearing for physical quantities, the remainder of this paper presents a study of meson masses containing at least one bottom quark.  Section \ref{sec:operators} explains how free-form smeared operators are built in lattice nonrelativistic QCD (NRQCD), using hydrogen-like wave-function shapes to target radial excitations and orbital excitations in addition to the ground state.  Section \ref{sec:SimDetails} provides details about the ensemble of configurations, about the NRQCD action for bottom quarks, and about the relativistic actions used for up/down, strange and charm quarks.

The masses and mass splittings obtained for the bottomonium spectrum are presented in Sec.~\ref{sec:bb}, including some states that have not yet been observed experimentally nor in previous lattice studies.  The $B$, $B_s$ and $B_c$ mesons are discussed in Sec.~\ref{sec:bx}, where the $B_c$ spectrum in particular provides predictions for upcoming experiments.  Previous lattice studies have considered some systematic effects that cannot be addressed in the present work and therefore provide a valuable context, so we make direct comparisons where possible.  In particular, lattice studies of the bottomonium spectrum can be found in Refs.~\cite{Davies:1994mp,Meinel:2009rd,Burch:2009az,Meinel:2010pv,Dowdall:2011wh,Dowdall:2013jqa,Daldrop:2011aa,Lewis:2012ir,Wurtz:2014uca} and lattice studies of bottom mesons appear in Refs.~\cite{Gregory:2010gm,McNeile:2012qf,Dowdall:2012ab,Allison:2004be,Gregory:2009hq,Burch:2015pka,Lang:2015hza,Bernardoni:2015nqa}.

\section{Free-Form Smearing}\label{sec:FFsmearing}

Quark smearing is a technique used frequently in lattice QCD spectroscopy to improve the signal obtained from correlation functions. The general idea is to smear the quark fields until they form an operator that resembles the wave function of the desired bound state. One popular method  \cite{Gusken:1989ad,Alexandrou:1990dq,Allton:1993wc} is to smear the quark field into a Gaussian shape by iteratively applying the discrete gauge-covariant Laplacian operator $\Delta$ as given by
\begin{align}
\tilde{\psi}(x)=\left(1+\frac{\alpha}{n}\Delta\right)^n\psi(x) \quad , \label{gaussiansmear}
\end{align}
where $n$ is the number of iterations and $\alpha$ is the smearing parameter. Note that the field is only smeared in the spatial directions and not the temporal direction. This approach is straightforward to implement, improves the ground-state signal, and suppresses excited states.

Another smearing method that is commonly used in lattice NRQCD is to fix the gauge links to Coulomb gauge and explicitly give the quark field an arbitrary shape \cite{Davies:1994mp} using the formula
\begin{align}
\tilde{\psi}(x)=\sum_y f(x-y)\,\psi(y) \quad . \label{gaugefixsmear}
\end{align}
One can tune the function $f(x-y)$ to obtain optimal signals not only for the ground state, but for excited states as well. By choosing a shape that closely matches the wave function of an excited state, the ground state can be suppressed and a cleaner excited-state signal obtained. The convolution can be efficiently implemented by a fast Fourier transform, and gauge fixing is required because Eq.~\eqref{gaugefixsmear} is not gauge invariant.

Free-form smearing \cite{vonHippel:2013yfa} combines the advantages of the previous methods. It allows the quark field to be smeared to an arbitrary shape while retaining gauge invariance, without the need for gauge fixing. Free-form smearing was initially applied to relativistic quarks, but we apply it for the first time to nonrelativistic heavy quarks. In our initial study of free-form smearing applied to bottomonium, a reduction in statistical errors relative to the gauge-fixed method was observed \cite{Wurtz:2014uca}.

In the original implementation of free-form smearing \cite{vonHippel:2013yfa}, the quark field
$\psi(x)$
at a single lattice site
$x$ 
was Gaussian smeared as in Eq.~\eqref{gaussiansmear} so that gauge links reach from site
$x$ 
to all other spatial sites on a given time slice. The free-form smeared field is given by the reweighting formula
\begin{align}
\tilde{\tilde{\psi}}(x)= \sum_yf(x-y)\frac{\tilde{\psi}_x(y)}{\left<\left\|\tilde{\psi}_x(y)\right\|\right>} \quad , \label{freeform}
\end{align}
where $f(x-y)$ is an arbitrary function,
$\tilde{\psi}_x(y)$ is the component of a Gaussian smeared field which gauge transforms at $x$ with the quark field $\psi$ located at $y$, and $\left<\left\|\tilde{\psi}_x(y)\right\|\right>$
is the ensemble average of the norm of the Gaussian smeared field. The norm is defined by
\begin{align}
\left\|\tilde{\psi}_x(y)\right\|=\sqrt{\operatorname{Tr}\left(\tilde{\psi}^\dag_x(y)\tilde{\psi}_x(y)\right)} \quad ,
\end{align}
where the trace is over spin and color indices. 
Note that the Gaussian smeared field $\tilde{\psi}(x)$ in Eq.~\eqref{gaussiansmear} is obtained from $\tilde{\psi}_x(y)$ by summing over $y$.
Any number of free-form smeared fields
$\tilde{\tilde{\psi}}(x)$
can be generated by reusing the same field
$\tilde{\psi}_x(y)$
and choosing a different shape $f(x-y)$.

Free-form smearing is fairly insensitive to the Gaussian smearing parameters $\alpha$ and $n$. The parameter $n$ should be chosen large enough so that gauge links reach every spatial lattice site, and $\alpha$ must be chosen so that $\frac{\alpha}{n}<\frac{1}{6}$, otherwise Gaussian smearing breaks down. For a lattice with $L$ sites in each of the spatial directions, $n=\frac{3}{2}L$ Gaussian smearing steps are required to reach all sites. This large number of Gaussian smearing steps is a computational burden for the original implementation of free-form smearing.

A much more computationally efficient alternative to the Gaussian version of free-form smearing is to replace the initial step (\ref{gaussiansmear}) by minimal paths, that is, the shortest paths along links from one lattice site to all other spatial sites. Recall that Gaussian smearing generates many link paths from one site to all other spatial sites, but that the links are also multiplied by a factor $\alpha/n$ which must be less than 1/6. Therefore, since the Gaussian method multiplies the links by a small factor, the shortest paths tend to dominate.

\begin{figure}[tb]
\centering
\vspace{5mm}
\includegraphics[scale=0.5]{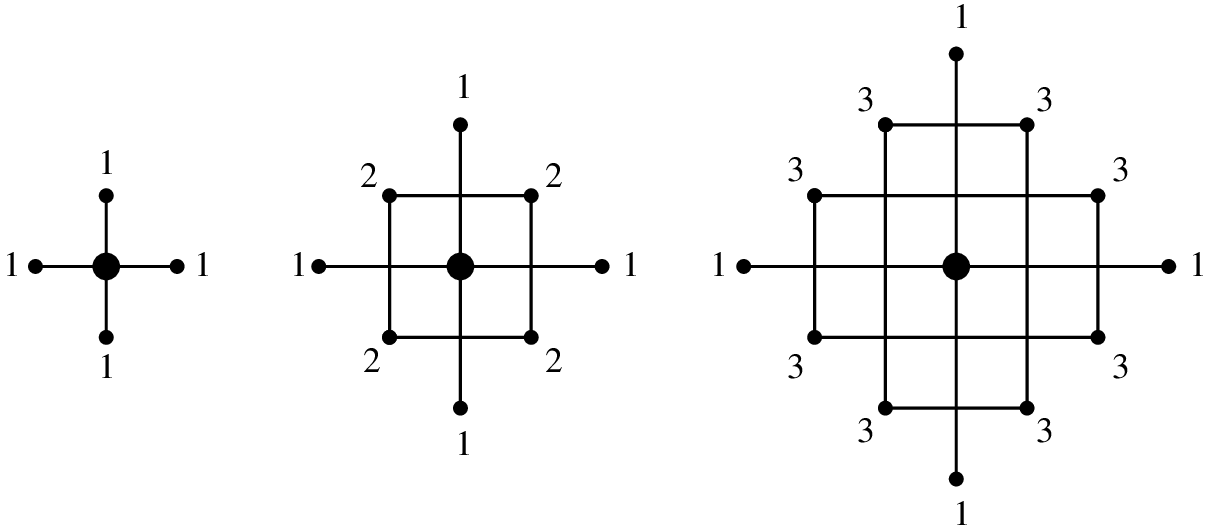}
\caption{Illustration of the minimal-path method of free-form smearing in two dimensions. The first, second and third iterations are shown from left to right, respectively. The source position is in the center of each sketch and links are multiplied outwards iteratively, forming a shell. Different link paths that come to the same site are summed; the number of minimal paths for each site is indicated.}
\label{figure-shortest_links}
\end{figure}

The new free-form link algorithm starts at a site
$x$
and multiplies links outward to all nearest neighbors, and then to all next-nearest outward neighbors, as illustrated in Fig.~\ref{figure-shortest_links}.
Paths along different links that lead to the same site are added together, thus resulting in a sum of minimal paths. This continues until all spatial sites have been reached and every link has been multiplied exactly once. The result is all minimal paths from the original site
$x$
to all other spatial sites
$y$,
\begin{align}
\tilde{\psi}_x(y)=\sum_\text{shortest paths}U(x\rightarrow y)\psi(y) \quad . \label{shortestpaths}
\end{align}
Free-form smearing can then be implemented as in Eq.~\eqref{freeform}. Whereas the Gaussian version of free-form smearing requires $O(L^4)$  link multiplications, the minimal-path method uses $3L^3$ link multiplications, which is exactly the number of spatial links on a given time slice.

Link smearing can also be used prior to the Gaussian smearing or minimal-path step. In the present work, stout-link smearing is used with parameters $\rho=0.15$ and $n_\rho=10$ as defined in Ref.~\cite{Morningstar:2003gk}. Therefore, our minimal-path method actually uses the shortest path of stout links, which effectively contains longer link paths as well. Link smearing is useful in reducing some excited-state contamination, although the most significant improvements for the present work are made by careful selection of the smearing shape $f(x-y)$.

\begin{figure}[tb]
\centering
\vspace{5mm}
\includegraphics[scale=0.5]{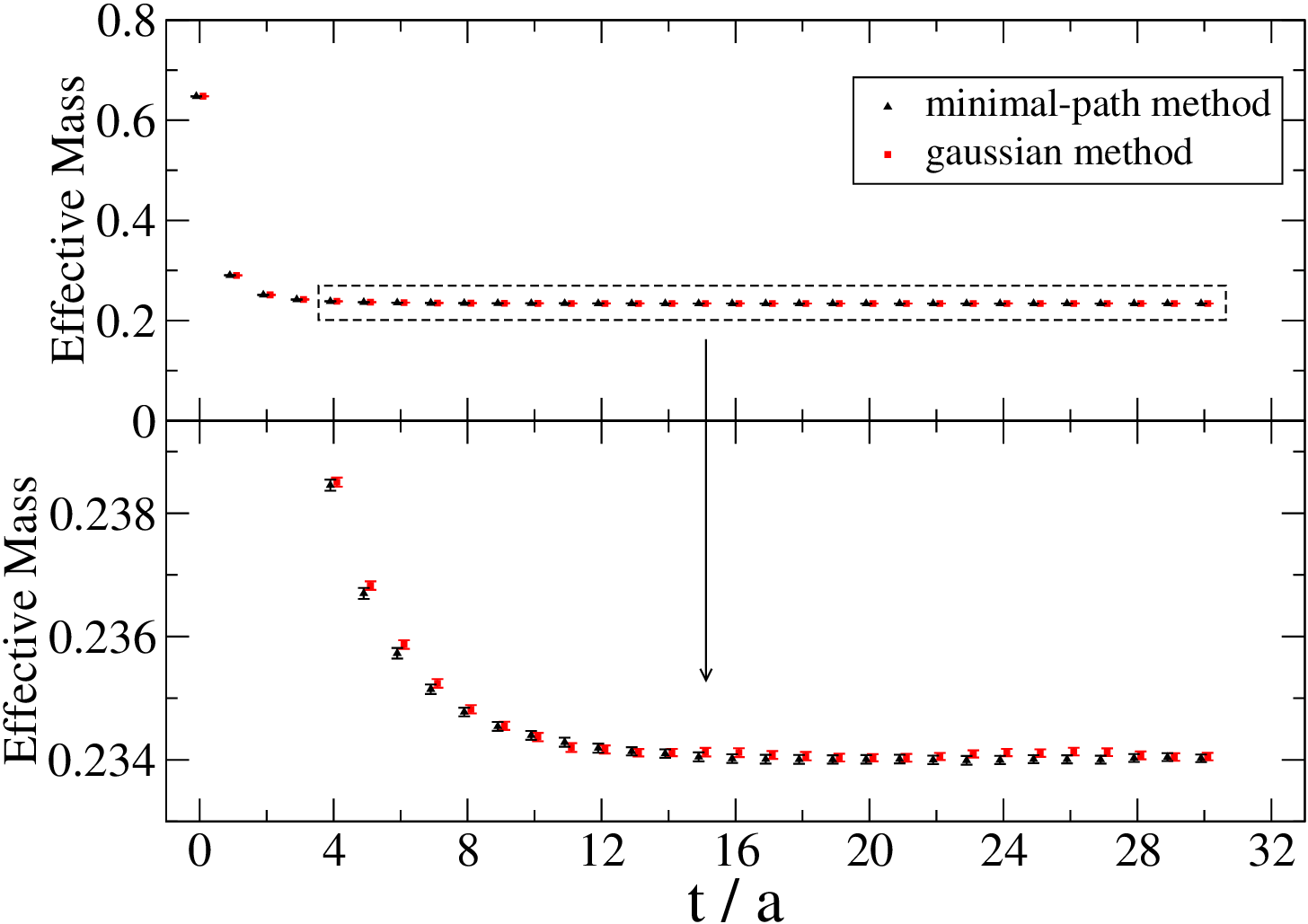}
\caption{Comparison of effective masses for the Gaussian and minimal-path versions of free-form smearing, using the same function $f(x-y)$, applied to a bottomonium pseudoscalar correlation function. The bottom panel shows a close-up view of the ground-state plateau. The minimal-path method is much more computationally efficient.}
\label{figure-effmass-bb-1S0}
\end{figure}

Figure~\ref{figure-effmass-bb-1S0} shows a direct comparison of the Gaussian and minimal-path free-form smearing methods for the effective mass of a bottomonium S-wave correlation function. The same wave-function shape $f(x-y)$ and stout-link smearings are used in both cases, and the Gaussian smearing parameters are $n=64$ and $\alpha=0.15$ from Eq.~\eqref{gaussiansmear}. There is virtually no difference between the results of the two methods shown in Fig.~\ref{figure-effmass-bb-1S0} and the same is true for other operators as well; what slight differences can be found (e.g. in a P-wave correlation function) are only an issue of fine-tuning in $f(x-y)$.

For the present study, smearing the field over the whole spatial volume is appropriate since the volume is not very much larger than the hadron size. This also conforms to the way Gaussian smearing, wave-function smearing and the original free-form smearing have been done in the past \cite{Gusken:1989ad,Alexandrou:1990dq,Allton:1993wc,vonHippel:2013yfa,Davies:1994mp}.

\section{Lattice Operators}\label{sec:operators}

For this study, smearing is applied to the bottom quark, leaving the antiquark unsmeared.  This is expedient because the bottom-quark propagator is obtained from lattice NRQCD and is less computationally expensive than the relativistic propagators for the up/down, strange and charm quarks.  For bottomonium, the resulting wave function places an anti-bottom quark at the origin surrounded by a bottom quark. For bottom mesons, an anti-up/down, anti-strange or anti-charm quark is surrounded by a bottom quark.  A physically intuitive picture would instead place the center of mass at the origin but, since we take a zero-momentum projection of the meson, the location of the center of mass is irrelevant; only the distance between the quark and antiquark matters.

Hydrogen-like (i.e. Coulomb potential) wave-function shapes have been used successfully within the gauge-fixed smearing method in lattice NRQCD\cite{Davies:1994mp} and we use them here with free-form smearing as well. Different shapes are used depending on the intended orbital angular momentum, and nodes are included to optimize the operator for radial excitations. Here is a list of the basic smearing shapes used in this paper:
\begin{align}
\text{S-wave:}& \quad f(x-y)=\left\{
\begin{array}{l}
e^{-\frac{r}{a_0}} \\
e^{-\frac{r}{a_0}}\,(r-b) \\
e^{-\frac{r}{a_0}}\,(r-c)(r-b)
\end{array} \right. \label{swave} \\
\text{P-wave:}& \quad f_i(x-y)=\left\{
\begin{array}{l}
e^{-\frac{r}{a_0}}\,\tilde{x}_i \\
e^{-\frac{r}{a_0}}\,\tilde{x}_i\,(r-b) 
\end{array} \right. \label{pwave} \\
\text{D-wave:}& \quad f_{ij}(x-y)=\left\{
\begin{array}{ll}
e^{-\frac{r}{a_0}}\,(\tilde{x}_i\tilde{x}_j-\frac{1}{3}\delta_{ij}(\tilde{x}_1^2+\tilde{x}_2^2+\tilde{x}_3^2)) \\
e^{-\frac{r}{a_0}}\,(\tilde{x}_i\tilde{x}_j-\frac{1}{3}\delta_{ij}(\tilde{x}_1^2+\tilde{x}_2^2+\tilde{x}_3^2))\, (r-b)
\end{array} \right. \label{dwave} \\
\text{F-wave:}& \quad f_{ijk}(x-y)=\tilde{x}_i \tilde{x}_j \tilde{x}_k\, e^{-\frac{r}{a_0}} \label{fwave} \\
\text{G-wave:}& \quad f_{ijkl}(x-y)=\tilde{x}_i \tilde{x}_j \tilde{x}_k \tilde{x}_l\, e^{-\frac{r}{a_0}} \quad , \label{gwave}
\end{align}
where $\tilde{x}_i$ is defined to be periodic, but $r$ is simply defined as the shortest distance between sites $x$ and $y$ in a periodic box:
\begin{align}
r=&\sqrt{(x_1-y_1)_\text{min}^2+(x_2-y_2)_\text{min}^2+(x_3-y_3)_\text{min}^2}\,,\\
\tilde{x}_i=&\sin\left(\frac{2\pi (x_i-y_i)}{L}\right) \quad .
\end{align}
The radius and nodal parameters $(a_0,b,c)$ are tuned to optimize the signal of the ground-state or radial excitations. Free-form smeared operators are built according to
\begin{align}\label{FullOperator}
\chi(x)\tilde{\tilde{\psi}}(x)=\chi(x)\sum_y\Omega_{f}(x-y)\frac{\tilde{\psi}_x(y)}{\left<\left\|\tilde{\psi}_x(y)\right\|\right>}
\end{align}
where $\psi$ is the quark, $\chi$ is the antiquark and $\Omega_{f}(x-y)$ is the free-form operator. Table~\ref{freeformoperators} lists all operators used in this paper by their lattice irreducible representations and continuum quantum numbers.
It provides a thorough coverage of quantum numbers for S, P and D waves plus five operators that offer the simplest exploration of some of the F and G waves.  As explained in Ref.~\cite{Lewis:2012ir}, additional F and G operators would duplicate some of the $\Lambda^{PC}$ quantum numbers that already exist in this table, so they are omitted here.
Sink operators are unsmeared and are given by the covariant derivative operators listed in Tables~II and III of Ref.~\cite{Lewis:2012ir}.
\begin{table}[tb]
\caption{The factor that defines the quantum numbers of the operator in Eq.~(\ref{FullOperator}) and incorporates the free-form smearing functions from Eqs.~\eqref{swave}--\eqref{gwave}.
Column 2 shows only the $J$ value that is expected to dominate in each case; for other possible $J$ values see Table I of Ref.~\cite{Lewis:2012ir}.}
\begin{tabular}{llll}
\hline\hline
$\Lambda^{PC}$ & $J^{PC}$ & $^{2S+1}L_J$ & $\Omega_{f}$ \\
\hline
$A_1^{-+}$&$0^{-+}$&$^1S_0$&$f$\\
$T_1^{--}$&$1^{--}$&$^3S_1$&$\{f\sigma_1,f\sigma_2,f\sigma_3\}$\\
$T_1^{+-}$&$1^{+-}$&$^1P_1$&$\{f_1,f_2,f_3\}$\\
$A_1^{++}$&$0^{++}$&$^3P_0$&$f_1\sigma_1+f_2\sigma_2+f_3\sigma_3$\\
$T_1^{++}$&$1^{++}$&$^3P_1$&$\{f_2\sigma_3-f_3\sigma_2\,f_3\sigma_1-f_1\sigma_3,f_1\sigma_2-f_2\sigma_1\}$\\
$E^{++}$  &$2^{++}$&$^3P_2$&$\{(f_1\sigma_1-f_2\sigma_2)/\sqrt{2},(f_1\sigma_1+f_2\sigma_2-2f_3\sigma_3)/\sqrt{6}\}$\\
$T_2^{++}$&$2^{++}$&$^3P_2$&$\{f_2\sigma_3+f_3\sigma_2,f_3\sigma_1+f_1\sigma_3,f_1\sigma_2+f_2\sigma_1\}$\\
$E^{-+}$  &$2^{-+}$&$^1D_2$&$\{(f_{11}-f_{22})/\sqrt{2},(f_{11}+f_{22}-2f_{33})/\sqrt{6}\}$\\
$T_2^{-+}$&$2^{-+}$&$^1D_2$&$\{f_{23},f_{31},f_{12}\}$\\
$T_1^{--}$&$1^{--}$&$^3D_1$&$\{f_{11}\sigma_1+f_{12}\sigma_2+f_{13}\sigma_3,f_{21}\sigma_2+f_{22}\sigma_2+f_{23}\sigma_3,f_{31}\sigma_1+f_{32}\sigma_2+f_{33}\sigma_3\}$\\
$E^{--}$  &$2^{--}$&$^3D_2$&$\{(f_{23}\sigma_1-f_{13}\sigma_2)/\sqrt{2},(f_{23}\sigma_1+f_{31}\sigma_2-2f_{12}\sigma_3)/\sqrt{6}\}$\\
$T_2^{--}$&$2^{--}$&$^3D_2$&$\{(f_{22}-f_{33})\sigma_1+f_{13}\sigma_3-f_{12}\sigma_2,(f_{33}-f_{11})\sigma_2+f_{21}\sigma_1-f_{23}\sigma_3,$\\
&&&$(f_{11}-f_{22})\sigma_3+f_{32}\sigma_2-f_{31}\sigma_1\}$\\
$A_2^{--}$&$3^{--}$&$^3D_3$&$f_{12}\sigma_3+f_{23}\sigma_1+f_{31}\sigma_2$\\
$T_1^{--}$&$3^{--}$&$^3D_3$&$\{3f_{11}\sigma_1-2f_{12}\sigma_2-2f_{13}\sigma_3,3f_{22}\sigma_2-2f_{23}\sigma_3-2f_{21}\sigma_1,$\\
&&&$3f_{33}\sigma_3-2f_{31}\sigma_1-2f_{32}\sigma_2\}$\\
$T_2^{--}$&$3^{--}$&$^3D_3$&$\{(f_{22}-f_{33})\sigma_1+2f_{12}\sigma_2-2f_{13}\sigma_3,(f_{33}-f_{11})\sigma_2+f_{23}\sigma_3-2f_{21}\sigma_1$,\\
&&&$(f_{11}-f_{22})\sigma_3+2f_{31}\sigma_1-2f_{32}\sigma_2\}$\\
$A_2^{+-}$&$3^{+-}$&$^1F_3$&$f_{123}$\\
$T_2^{+-}$&$3^{+-}$&
$^1F_3$
&$\{f_{122}-f_{133},f_{233}-f_{211},f_{311}-f_{322}\}$\\
$A_2^{++}$&$3^{++}$&$^3F_3$&$(f_{221}-f_{331})\sigma_1+(f_{332}-f_{112})\sigma_2+(f_{113}-f_{223})\sigma_3$\\
$T_1^{-+}$&$4^{-+}$&$^1G_4$&$\{f_{2223}-f_{3332},f_{3331}-f_{1113},f_{1112}-f_{2221}\}$\\
$A_1^{--}$&$4^{--}$&$^3G_4$&$(f_{2223}-f_{3332})\sigma_1+(f_{3331}-f_{1113})\sigma_2+(f_{1112}-f_{2221})\sigma_3$\\
\hline\hline
\end{tabular}
\label{freeformoperators}
\end{table}

\begin{figure}[tb]
\centering
\vspace{5mm}
\includegraphics[scale=0.5]{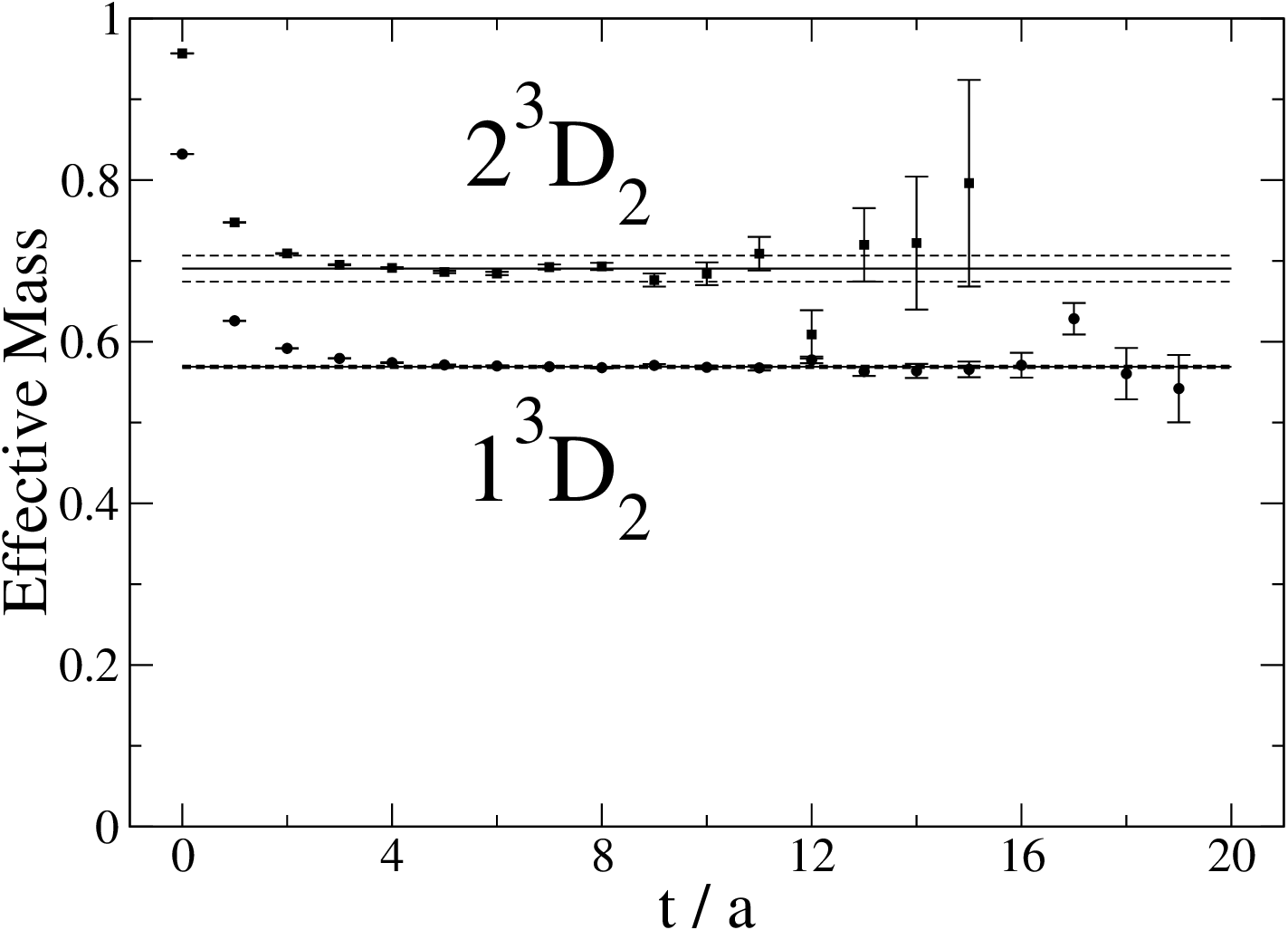}
\caption{Effective mass plots for the free-form smeared $^3D_2$ ($E^{--}$ representation) bottomonium correlation functions tuned to optimize the ground state and the first excited state. Energy values extracted from fits to the correlation functions are shown as solid horizontal lines. Dashed horizontal lines are the statistical uncertainties.}
\label{figure-effmass-bb-3D2E}
\end{figure}

\begin{figure}[tb]
\centering
\vspace{5mm}
\includegraphics[scale=0.5]{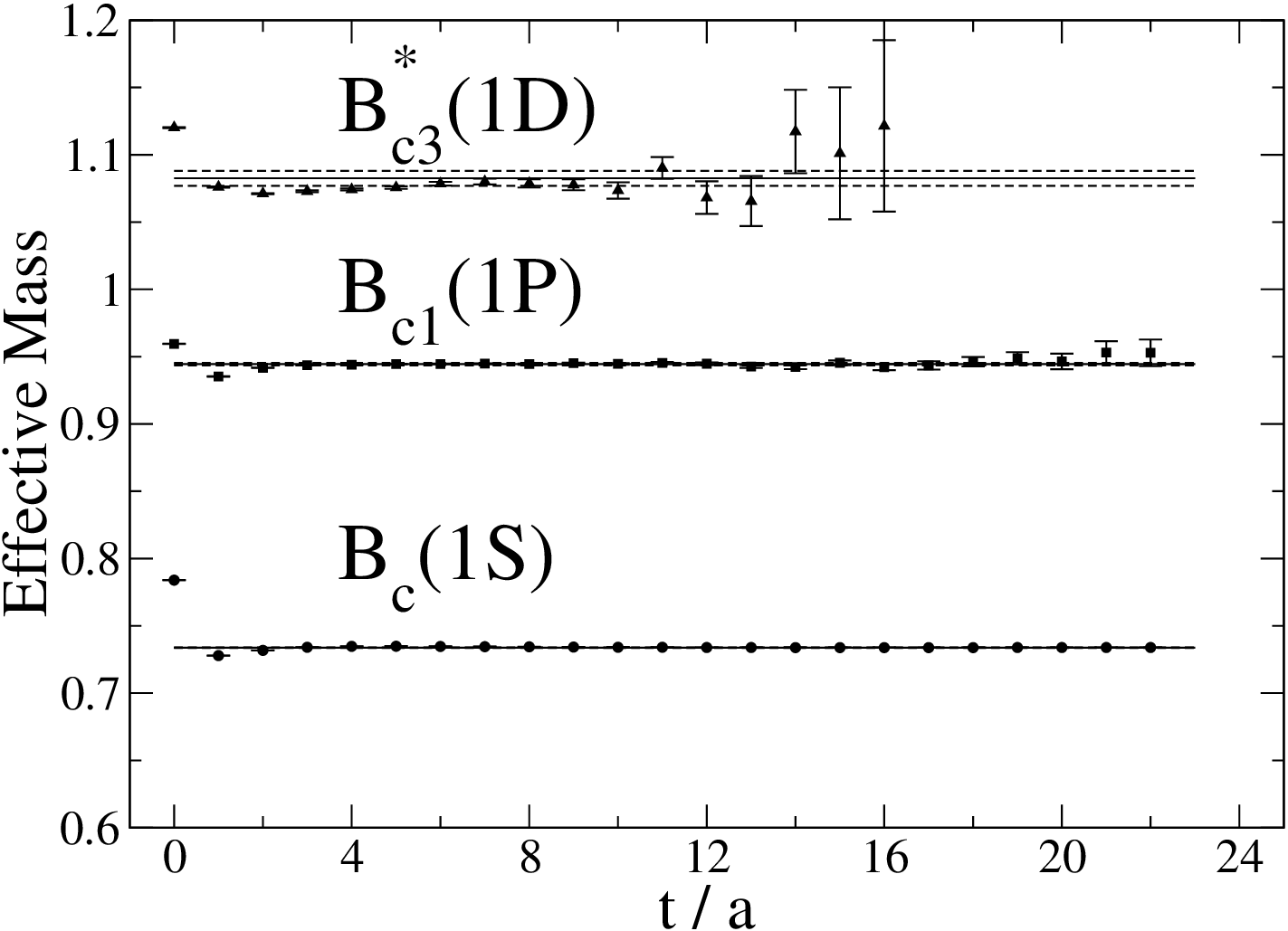}
\caption{Effective mass plots for the free-form smeared $0^-$, $1^+$ and $3^-$ ($T_2^-$ representation) $B_c$ meson correlation functions tuned to optimize the S-wave, P-wave and D-wave ground states. The P-wave operator shown corresponds to $1^{++}$ when applied to quarkonium; see Sec.~\ref{sec:bx} for discussion on this operator when applied to heavy-light mesons. Energy values extracted from fits to the correlation functions are shown as solid horizontal lines. Dashed horizontal lines are the statistical uncertainties.}
\label{figure-effmass-Bc-1S1P1D}
\end{figure}

The optimized free-form smearing parameters corresponding to Eqs.~\eqref{swave}--\eqref{gwave} used in this study are given in Table~\ref{freeformparameters}.
\begin{table}[tb]
\caption{Free-form smearing parameters for $B$, $B_s$, $B_c$ and bottomonium. Parameters were optimized by trial and error. In addition, nonoptimal parameters ($a_0=0.5$ and 1.0) are used for $B$, $B_s$ and $B_c$, as well as unsmeared operators for bottomonium, as discussed in the text.}
\begin{tabular}{llllllllllllllll}
\hline\hline
\multicolumn{4}{c}{bottomonium}&&\multicolumn{3}{c}{$B_c$}&&\multicolumn{3}{c}{$B_s$}&&\multicolumn{3}{c}{$B$} \\
\cline{1-4}\cline{6-8}\cline{10-12}\cline{14-16}
$L\quad$ & $a_0\quad$ & $b\quad$ & $c\quad$ &$\quad$& $L\quad$ & $a_0\quad$ & $b\quad$ &$\quad$& $L\quad$ & $a_0\quad$ & $b\quad$ &$\quad$& $L\quad$ & $a_0\quad$ & $b\quad$ \\
\hline
$S$&1.4&&         &&$S$&0.5&     &&$S$&0.5&     &&$S$&0.5&\\
   &1.6&&         &&   &2.0&     &&   &4.5&     &&   &5.0&\\
   &2.6&2.6&      &&   &2.2&     &&   &5.8&5.8  &&   &5.5&\\
   &2.8&2.8&      &&   &4.0&4.0  &&   &6.0&6.0  &&   &6.2&6.2\\
   &3.0&2.13&6.0  &&$P$&1.0&     &&$P$&1.0&     &&$P$&0.5&\\
$P$&2.0&&         &&   &3.0&     &&   &6.0&     &&   &7.0&\\
   &3.0&4.6&      &&   &4.0&6.0  &&   &7.0&     &&   &&\\
$D$&2.5&&         &&$D$&1.0&     &&   &&        &&   &&\\
   &3.5&6.5&      &&   &5.0&     &&   &&        &&   &&\\
$F$&3.0&&         &&&&           &&   &&        &&\\
   &4.0&&         &&\\
$G$&4.0&&         &&\\
   &5.0&&         &&\\
\hline\hline
\end{tabular}
\label{freeformparameters}
\end{table}
The parameters $(a_0,b,c)$ are tuned by hand to optimize the signal in a correlation function for the ground state, the first excited state and, for bottomonium S waves, the second excited state. These optimizations improve the accuracy and precision of the spectrum.
The effective mass plots shown in Figs.~\ref{figure-effmass-bb-3D2E} and \ref{figure-effmass-Bc-1S1P1D} demonstrate the effectiveness of free-form smearing to obtain clean signals for ground states and excited states. Figure~\ref{figure-effmass-bb-3D2E} shows that the plateau begins at small Euclidean times for a bottomonium D-wave ground state and for the first radial excitation. 
In particular, note that the effective mass for the first radial excitation shows no contamination from the ground state. 
Similarly, Fig.~\ref{figure-effmass-Bc-1S1P1D} shows small-time plateaus for $B_c$ meson S-wave, P-wave and D-wave ground states.

Free-form smearing shrinks the contamination from unwanted states, thereby emphasizing the one state of interest, but it does not eliminate the contamination entirely.  Different choices for the smearing parameters can be used to emphasize each state separately, and then a simultaneous fit to several correlation functions will be stable and reliable.  For the present work, nonoptimized correlation functions are also included in the multistate fits and they further stabilize the results.  For bottomonium, the nonoptimized correlators are calculated using unsmeared operators as given in Tables~II and III of \cite{Lewis:2012ir}. For $B$, $B_s$ and $B_c$ ground-state profiles, small radial parameters $a_0=0.5$ or $a_0=1.0$ are used as the nonoptimized choices.

Chronologically, the calculation for $B_c$ was started before we had developed the minimal-path method, so it uses the Gaussian version of free-form smearing with parameters $n=64$ and $\alpha=0.15$ from Eq.~\eqref{gaussiansmear}.  The minimal-path method of free-form smearing is used for all $B$, $B_s$ and bottomonium calculations.  Throughout this work, stout links with parameters $\rho=0.15$ and $n_\rho=10$ were employed.

\section{Simulation Details}\label{sec:SimDetails}

An $N_f=2+1$ gauge field ensemble from the PACS-CS Collaboration \cite{Aoki:2008sm} was used for this study. The action contains an Iwasaki gauge term and clover-improved Wilson fermions. The lattice dimensions are $32^3\times64$ and the lattice spacing
\begin{equation}\label{aORIG}
a_{\rm PACS-CS}=0.0907(13) {\rm ~fm}
\end{equation}
was determined by the PACS-CS Collaboration \cite{Aoki:2008sm} using the experimental $\pi$, $K$ and $\Omega$ masses. The pion mass is near physical at $m_\pi=156(7)$ MeV. At $(m_K)_{\rm orig}=554(8)$ MeV, the kaon mass is larger than its physical value so, to account for this, a partially quenched strange quark is used in this study as was done in Ref.~\cite{Lang:2014yfa} where the valence strange quark was retuned to match the physical $\phi$ mass, giving a kaon mass of $m_K=504(7)~\text{MeV}$.

The charm quark parameters are taken from Ref.~\cite{Lang:2014yfa} where they were tuned using the Fermilab interpretation. The parameters for the relativistic quarks are given in Table~\ref{quarkparameters}.
\begin{table}[tb]
\caption{The hopping parameters and clover coefficients for relativistic sea and valence quarks.}
\label{quarkparameters}
\begin{tabular}{lll}
\hline \hline
Quark flavor  & $\kappa$ & $c_{SW}$ \\
\hline
sea quarks    & \\
up/down       & 0.13781  & 1.715    \\
strange       & 0.13640  & 1.715    \\
valence quarks& \\
up/down       & 0.13781  & 1.715    \\
strange       & 0.13666  & 1.715    \\
charm         & 0.12686  & 1.64978  \\
\hline \hline
\end{tabular}
\end{table}
The propagators for the relativistic valence quarks ($u/d,s,c$) are calculated using the sap\_gcr solver from the DD-HMC code made available by Martin L\"uscher \cite{Luscher:2003qa,Luscher:2005rx,ddhmc}.

The bottom quark is implemented using lattice NRQCD, including terms up to $O(v^4)$ in the bottom-quark velocity, which corresponds to $c_i=1$ for $1\leq i\leq6$ and $c_i=0$ for $i\geq 7$ according to the notation of the Appendix in Ref.~\cite{Lewis:2008fu}. Following Ref.~\cite{Lewis:2012ir}, the bare mass of the bottom quark ($M_b=1.95$) is taken from fitting the kinetic mass of the $\Upsilon$ to its experimental value, the tadpole factor is set to the average link in Landau gauge ($u_L=0.8463$), and the stability parameter is chosen to be $n=4$.

The calculation of correlation functions for bottom mesons with a nonrelativistic bottom quark requires some additional care with regard to the contraction of spin indices. Our choice for Dirac $\gamma$ matrices is
\begin{align}
\gamma_k=\left(\begin{array}{cc}
0        & \sigma_k \\
\sigma_k & 0        \end{array}\right) \quad\quad
\gamma_4=\left(\begin{array}{cc}
I & 0  \\
0 & -I \end{array}\right) \quad ,
\end{align}
which allows the Dirac indices of the relativistic propagator and the Pauli indices of the nonrelativistic propagator to be contracted in a simple manner. For forward time propagation the top two Dirac indices are used, while for backward time propagation the bottom two Dirac indices are used.

Simulating multiple sources on a given configuration can reduce statistical errors. A random $U(1)$ wall source imitates multiple sources at different spatial sites on a given time slice without the need to calculate many propagators. While it is easily implemented using the conventional smearing methods given by Eqs.~\eqref{gaussiansmear} and \eqref{gaugefixsmear}, a free-form smeared random wall source is computationally expensive because every site
$x$
must be smeared independently and summed according to the formula
\begin{align}
\tilde{\tilde{\psi}}_W=\sum_i^N e^{i\theta_W(x_i)}\, \tilde{\tilde{\psi}}(x_i)
\end{align}
where $0\leq\theta_W(x_i)<2\pi$ is a random complex phase at each lattice site.
An efficient approach is to calculate a partial random wall source, where only a relatively small number of spatial sites
$x$
is used. For this study the partial wall source has $N=4^3$ evenly spaced free-form smeared sites.
With the original implementation of free-form smearing, the cost of free-form smearing this partial wall source was comparable to the cost of calculating one NRQCD propagator over the full time extent of the lattice.  The new implementation of free-form smearing reduced the cost by a factor near 100.
Including more sites in the partial wall source did not improve the statistical errors for the case of bottomonium \cite{Wurtz:2014uca}. Statistical errors are further reduced by using partial wall sources for multiple time slices. For bottomonium a partial wall source is built for each of the 64 time slices, and for the $B$, $B_s$ and $B_c$ mesons a partial wall source is built for 32 time slices. To further increase the statistics, NRQCD propagators are calculated forward and backward in time. The correlation functions are binned over all time sources and over forward and backward propagation.

The simulation energies are extracted by a simultaneous multiexponential fit to multiple free-form smeared correlation functions of the form
\begin{align}
C^i(t)=\sum_n A_n^i\, e^{-E_n t} \quad , \label{expfit}
\end{align}
where $E_n$ is the fit parameter for the energy of the $n\text{th}$ state. Each $\Lambda^{PC}$ channel from Table  \ref{freeformoperators} is fit separately. The fits are done to correlation functions with free-form smeared source operators as described in Eqs.~\eqref{swave}--\eqref{gwave} and Tables~\ref{freeformoperators} and \ref{freeformparameters}. For bottomonium, unsmeared source operators are used as well. All fits exclude the source time step at $t=0$, and are typically truncated at $t=15$ or $t=23$ because the signal is lost in noise for larger Euclidean times.

\section{Bottomonium Spectrum}\label{sec:bb}

For bottomonium, the S-wave correlator is fitted to six exponentials, where the lightest three energies are identified as the ground state, first radial excitation and second radial excitation. The other exponentials are collectively able to account for higher excitations but we do not interpret them individually as the physical states.

For the P waves and D waves, separate fits are performed to extract the ground states and first excited states. 
The fits for the P-wave and D-wave excited states are done using three correlation functions: an unsmeared operator, a free-form smeared ground state, and a free-form smeared excited state. The ground-state P waves and D waves are fit to the first two correlation functions with the excited-state optimized correlator excluded. Removing this correlation function from the fit significantly reduced the statistical uncertainty of the ground states. These more precise ground-state fit values are statistically compatible with the noisier ground-state fit values that are obtained by including the excited-state correlation function. 
The excited-state fits use five exponentials (six for $^3D_1$ and $^3D_{3T_1}$) and the ground-state fits use four (five for $^3D_1$ and $^3D_{3T_1}$). The D-wave $T_1^{--}$ operators $^3D_1$ and $^3D_{3T_1}$ have a noticeable mixing with the $^3S_1$ ground state and an additional exponential is included in the fit for these two channels. The smallest energy is consistent with the $^3S_1$ ground state and all higher energy levels are assumed to be D waves. A confident interpretation would require a cross-correlation matrix between the $^3S_1$, $^3D_1$ and $^3D_{3T_1}$ operators, but this is beyond the scope of the present study. We note that the results for $^3S_1$ also assume no D-wave contamination.

The F-wave bottomonium state $^3F_3$ is fit with four exponentials to three correlation functions: unsmeared, free-form smeared with $a_0=3.0$, and free-form smeared with $a_0=4.0$. To obtain a reliable result, $^1F_{3A_2}$ is fit with four exponentials to the unsmeared and $a_0=3.0$ free-form smeared correlation functions, and $^1F_{3T_2}$ is fit with three exponentials to the $a_0=3.0$ and $a_0=4.0$ free-form smeared correlation functions. The bottomonium G-wave ground states are obtained from four exponential fits to three correlations functions: unsmeared, free-form smeared with $a_0=4.0$ and free-form smeared with $a_0=5.0$. Even for these higher orbital angular momenta, the hydrogen-like smearing profiles from Eqs.~\eqref{fwave} and \eqref{gwave} produced significantly better ground-state signals than Gaussian smearing.
\begin{table}[tb]
\caption{Bottomonium mass splittings for $\overline{1P}$, $\overline{2S}$, $1^3D_2$ and $\overline{2P}$ with respect to $\overline{1S}$, where a bar represents the spin average. Results using two different lattice spacing definitions are shown: Eq.~\eqref{aORIG} and Eq.~\eqref{newlatticespacing}. The value for $^3D_2$ is the dimensional average of the $E$ and $T_2$ lattice representations.}
\begin{tabular}{llll}
\hline\hline
~~~~~~~~~ & \multicolumn{3}{c}{$m-m_{\overline{1S}}$ [MeV/$c^2$]} \\
\cline{2-4}
& lattice using $a_\text{PACS-CS}$~~~ & lattice using $a_{\overline{1P}-\overline{1S}}$~~~ & experiment \cite{PDG} \\
\hline
$\overline{1P}$    & 437(6)  & 455.0(9) & 455.0(9)   \\
$\overline{2S}$    & 547(10) & 569(6)   & 572.5(1.3) \\
$1^3D_{2}$         & 688(10) & 715(3)   & 719.0(1.7) \\
$\overline{2P}$    & 743(15) & 773(11)  & 815.4(9)   \\
\hline\hline
\end{tabular}
\label{table-differentlatticespacing}
\end{table}

Since the bottom-quark mass is fixed by tuning the kinetic mass of the $\Upsilon$ to its physical value \cite{Lewis:2012ir}, the absolute masses of the bottomonium spectrum are calculated from
\begin{align}\label{offset}
mc^2=m_\Upsilon^\text{exp}c^2+\frac{\hbar c}{a}\left(E^\text{sim}-E_\Upsilon^\text{sim}\right) \quad .
\end{align}
Using the lattice spacing scale given in Ref.~\cite{Aoki:2008sm}, the masses of the $2S$, $1P$, $2P$ and $1D$ bottomonia are all systematically smaller than the experimental values by a significant amount, as shown in Table~\ref{table-differentlatticespacing}.
However, ratios of differences having the $1S$ mass subtracted from the $2S$, $1P$ and $1D$ masses agree with experiment, which suggests using the bottomonium spectrum to set the scale.
A new lattice spacing is defined using the spin-averaged $\overline{1P}-\overline{1S}$ mass splitting
\begin{align}
a_{\overline{1P}-\overline{1S}}=\hbar c\frac{\left(E_{\overline{1P}}-E_{\overline{1S}}\right)_\text{sim}}{\left(m_{\overline{1P}}c^2-m_{\overline{1S}}c^2\right)_\text{exp}}=0.0872(3)\text{fm}\quad,\label{newlatticespacing}
\end{align}
which is 4.0\% smaller than the PACS-CS value from Eq.~(\ref{aORIG}). Reference~\cite{Lang:2014yfa} noted that other methods find a lattice spacing that is up to 4.4\% larger than the PACS-CS value for this ensemble.

The lattice spacing reported in Ref.~\cite{Aoki:2008sm} was obtained from the light quark hadron spectrum, which is less relevant for the case of bottomonium.
Therefore, we use Eq.~\eqref{newlatticespacing} to set the scale for the bottomonium spectrum.
Even after using the scale from Eq.~\eqref{newlatticespacing}, the $2P$ masses remain systematically smaller than experiment.
This cannot be due to contamination from higher excited states because that would cause the $2P$ masses to be larger, not smaller.
Applying radiative corrections or nonperturbative tunings to the NRQCD coefficients $c_i$ and including higher-order terms in the bottomonium velocity are possible ways to remove this discrepancy.
In principle, the bottom-quark mass should also be retuned using this new lattice spacing.
At present, we simply note that setting the scale with physics relevant to the bottom quark increases the accuracy of the bottomonium spectrum.

\begin{figure}[tb]
\centering
\includegraphics[scale=0.6,clip=true]{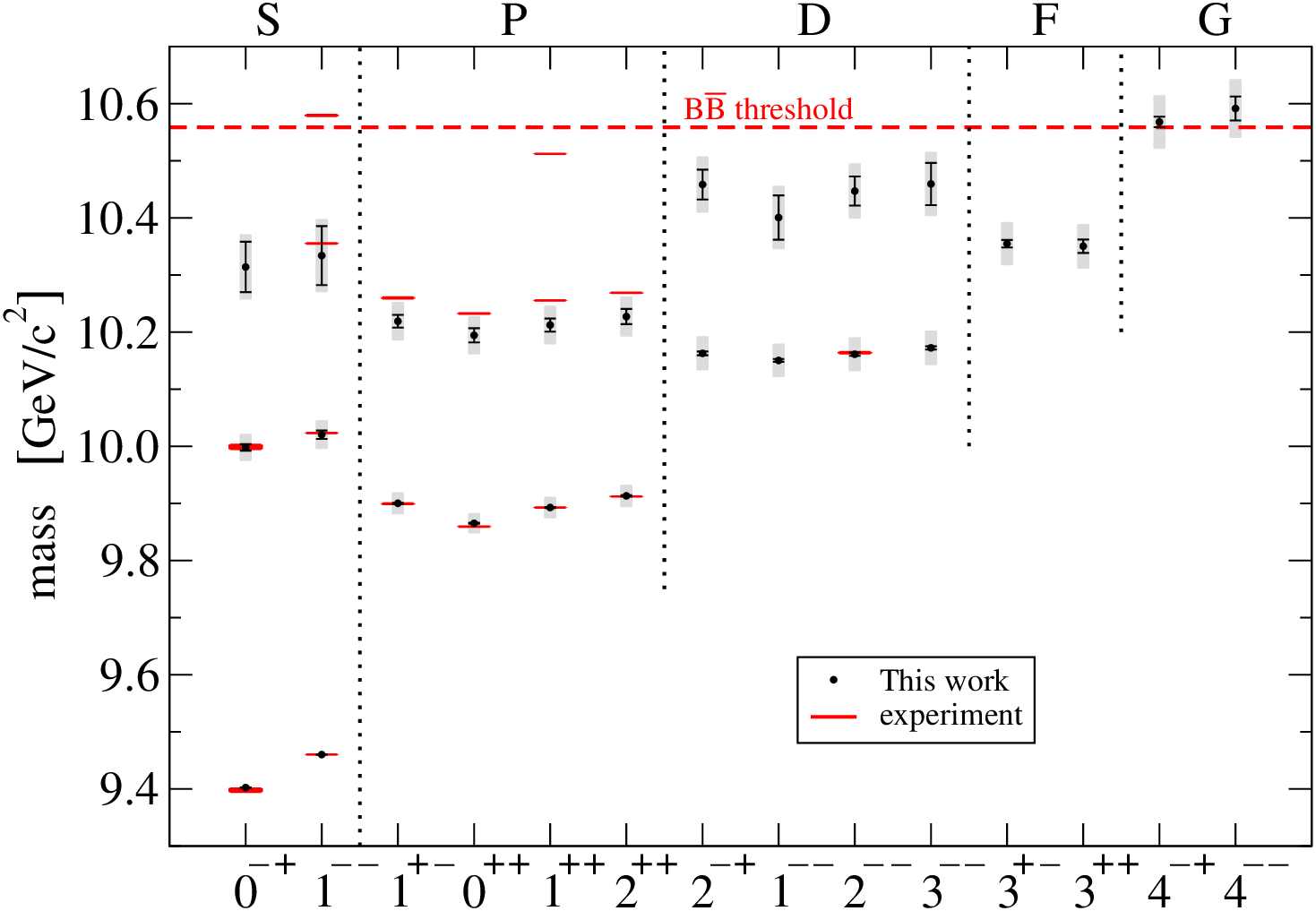}
\caption{Mass spectrum of bottomonium. Red bands are experimental values. Black points with errors bars are lattice data with statistical errors only. Grey bands are the statistical error and lattice spacing uncertainty, added in quadrature. Numerical values are given in Table~\ref{table-spectrum-bottomonium}.}
\label{bottomonium-spectrum}
\end{figure}

The entire bottomonium spectrum below the $B\overline{B}$ threshold [with the exception of the $3P$ states, where the experimental value of $\chi_{b1}(3P)$ is just below the $B\overline{B}$ threshold \cite{Aaij:2014hla}] is shown in Fig.~\ref{bottomonium-spectrum}, as extracted using chi-squared fits of free-form smeared correlation functions.
For spin-2 and spin-3 states, where results for more than one lattice irreducible representation $\Lambda$ were calculated, the dimensional average of the simulation energies,
\begin{align}
E_{\overline{\text{dim}}} = \frac{\sum_\Lambda \text{dim}(\Lambda)E_\Lambda}{\sum_\Lambda\text{dim}(\Lambda)} \quad,
\end{align}
is our reported value for the mass.
The grey bands show a combination of the statistical errors and a 4.0\% systematic uncertainty in the lattice spacing, which come from discrepancies in the determination of the lattice spacing.
Our work provides the first lattice result for the bottomonium D-wave radial excitations in all channels.

Precise spin splittings were obtained for $1S$, $2S$, $1P$, $2P$ and $1D$ bottomonia, shown in Table~\ref{table-bbsplittings}. The spin splittings in Table~\ref{table-bbsplittings} agree well with experiment, except for the $\chi_{b0}(1P)$ which is larger. This discrepancy with the BABAR results \cite{Lees:2014qea} is more evident in the spin-dependent splittings
\begin{align}
-2\chi_{b0}(1P)+3\chi_{b1}(1P)-\chi_{b2}(1P) &= \left\{\begin{array}{ll}
34.5(9)\,\text{MeV}\quad   & \text{this work} \\
46.0(1.9)\,\text{MeV}\quad & \text{BABAR}
\end{array}\right. \label{pwave-splitting-c4} \\
-2\chi_{b0}(1P)-3\chi_{b1}(1P)+5\chi_{b2}(1P) &= \left\{\begin{array}{ll}
157(4)\,\text{MeV}     \quad & \text{this work} \\
160.0(2.2)\,\text{MeV} \quad & \text{BABAR}
\end{array}\right. \label{pwave-splitting-c3}
\end{align}
where, at tree level, the former is proportional to the NRQCD parameter $c_4^2$ while the latter is proportional to $c_3$. For $\chi_{b2}(1P)$ in Eqs.~\eqref{pwave-splitting-c4} and \eqref{pwave-splitting-c3} the dimensional average of the $E$ and $T_2$ lattice irreducible representations is used. Since our simulations only use tree-level coefficients where $c_3=c_4=1$, the P-wave spin structure could be improved by tuning $c_4$ until Eq.~\eqref{pwave-splitting-c4} agrees with experiment. Reference~\cite{Dowdall:2011wh} noted that increasing $c_4>1$ had the effect of decreasing the $\chi_{b0}(1P)$ mass relative to the $\overline{1^3P}$ spin-averaged mass while doing little else.

\begin{table}[tb]
\caption{Spin splittings for $1S$, $2S$, $1P$, $2P$ and $1D$ bottomonia. $\overline{^3P}$ represents the spin-averaged triplet P wave. Only statistical errors are shown for this work. \\$^a$ Extracted from Fig.~5 of Meinel \cite{Meinel:2009rd} and Fig.~23 of HPQCD \cite{Dowdall:2011wh}. \\$^b$ Extracted from Ref.~\cite{Daldrop:2011aa}.}
\begin{tabular}{lllll}
\hline\hline
                                      & this work & Meinel \cite{Meinel:2009rd,Meinel:2010pv} & HPQCD \cite{Dowdall:2013jqa,Dowdall:2011wh,Daldrop:2011aa} & experiment \cite{PDG} \\
\hline
$\Upsilon(1S)-\eta_b(1S)$             & 57.97(16) & 60.3(5.5)(5.0)(2.1) & 62.8(6.7)         & 62.3(3.2)~~[Belle=57.9(2.3)] \\
$\Upsilon(2S)-\eta_b(2S)$             & 22(3)     & 23.5(4.1)(2.1)(0.8) & 26.5(1.6)(1.4)    & 24(4)     \\
$\overline{1^3P}-\chi_{b0}(1P)$       & 35.7(5)   & 36(3)$^a$        & 40.0(1.2)$^a$  & 40.4(5)   \\
$\overline{1^3P}-\chi_{b1}(1P)$       & 8.3(4)    & 8.5(1.8)$^a$     & 7.4(6)$^a$     & 7.1(4)    \\
$\overline{1^3P}-h_b(1P)$             & 0.8(3)    & 0.04(93)(20)        & 2.0(6)$^a$     & 0.6(1.0)  \\
$\chi_{b2}(1P)-\overline{1^3P}$       & 12.1(3)   & 12.4(1.4)$^a$    & 12.4(5)$^a$    & 12.3(2)   \\
$\overline{2^3P}-\chi_{b0}(2P)$       & 24(9)     & -                   & -                 & 27.7(7)   \\
$\overline{2^3P}-\chi_{b1}(2P)$       & 6(5)      & -                   & -                 & 4.8(5)    \\
$\overline{2^3P}-h_b(2P)$             & 0(4)      & -                   & -                 & 0.4(1.3)  \\
$\chi_{b2}(2P)-\overline{2^3P}$       & 9(4)      & -                   & -                 & 8.4(3)    \\
$\Upsilon_2(1D)-\Upsilon  (1D)$       & 11(3)     & -                   & 17(6)             & -         \\
$\Upsilon_3(1D)-\Upsilon_2(1D)$       & 11(2)     & -                   & 18(5)             & -         \\
$\Upsilon_3(1D)-\Upsilon  (1D)$       & 22(3)     & -                   & 34(8)             & -         \\
$\eta_{b2}(1D)- \Upsilon_2(1D)$       & 1.6(1.5)  & -                   & 5(8)$^b$          & -         \\
\hline\hline
\end{tabular}
\label{table-bbsplittings}
\end{table}

The D-wave splittings have not yet been experimentally observed, but our results in Table~\ref{table-bbsplittings} are smaller than the predictions given in Ref.~\cite{Daldrop:2011aa}, also shown in Table~\ref{table-bbsplittings}. Reference~\cite{Daldrop:2011aa} proposed a method to reduce systematic effects from $c_3$ and $c_4$ dependence. When applied to our data, this method produces results that are consistent within statistical uncertainties with the simple difference shown in Table~\ref{table-bbsplittings}.

\section{$B$, $B_s$ and $B_c$ Spectrum}\label{sec:bx}

In contrast to the case of bottomonium, charge conjugation is not a helpful quantum number for bottom mesons.  Whereas the $1^{+-}$ and $1^{++}$ operators in Table~\ref{freeformoperators} couple to separate quarkonium states, those same two operators each couple to a mixture of heavy-light meson states \cite{Boyle:1997rk,Bali:2003jv}.  The same is true for the $2^{-+}$ and $2^{--}$ operators.

In the nonrelativistic basis, the $1^{+-}$ and $1^{++}$ operators are distinguished by the presence of a Pauli matrix in the latter but not in the former.  Terms in the
NRQCD
propagator that contain a Pauli matrix are also proportional to an odd power of the quark momentum. The cross correlator of a Pauli matrix operator with a non-Pauli matrix operator will be proportional to an odd power of momentum, and should be zero in the ensemble average because of spatial lattice symmetry. We verified that cross correlators of the $1^{+-}$ and $1^{++}$ operators for bottom mesons are statistically consistent with zero at all Euclidean times. This confirms that the two operators are orthogonal, but it does not provide information on the mixing of the physical states within each operator. The same is true for the $2^{-+}$ and $2^{--}$ operators.

In practice the $1^{+-}$ and $1^{++}$ operators appear to plateau at different energies, given a limited extent in Euclidean time. Even though they should contain the same ground state, they have a mixing with an excited state that is very close in energy to the ground state. They each have a different mixing with these two states, which can give the appearance of different ground states in a practical lattice study. These false ground-state signals would each be larger than the physical ground state and smaller than the physical excited state. A multiexponential fit is unable to distinguish the two physical states given the precision of our data. An application of the variational method to a correlator matrix can separate these states (see, for example, Ref.~\cite{Mohler:2011ke}), but that is beyond the scope of this project. We will state the results from both $1^+$ operators (and both $2^-$ operators), using a prime for the heavier of the pair, and acknowledge that there is an unresolved mixing.

For $B_c$ mesons, the S-wave correlator is fitted with five exponentials, while the P and D waves use four exponentials. The D-wave $T_1^-$ operators have an overlap with the S-wave $B_c^*$. The smallest energy is interpreted as the $B_c^*$ ground state and next smallest energy as the D-wave ground state. For $B_s$ and $B$, the S- and P-wave correlators are fitted with four exponentials.

For bottom mesons, the additive NRQCD mass of Eq.~(\ref{offset}) takes the form
\begin{align}
mc^2=\frac{1}{2}m_\Upsilon^\text{exp}c^2+\frac{\hbar c}{a}\left(E^\text{sim}-\frac{1}{2}E_\Upsilon^\text{sim}\right)
\end{align}
and the lattice spacing from Eq.~\eqref{newlatticespacing} is used. Absolute masses are not calculated for $B_c$ mesons. The absolute mass for a $B_c$ meson contains large discretization effects because the mass of the charm quark is large compared to our lattice cutoff, so absolute masses will not be studied in this work. Instead, we focus on mass differences among $B_c$ states because these are expected to be close to their physical values \cite{Lang:2014yfa}.

\begin{figure}[tb]
\centering
\includegraphics[scale=0.6,clip=true]{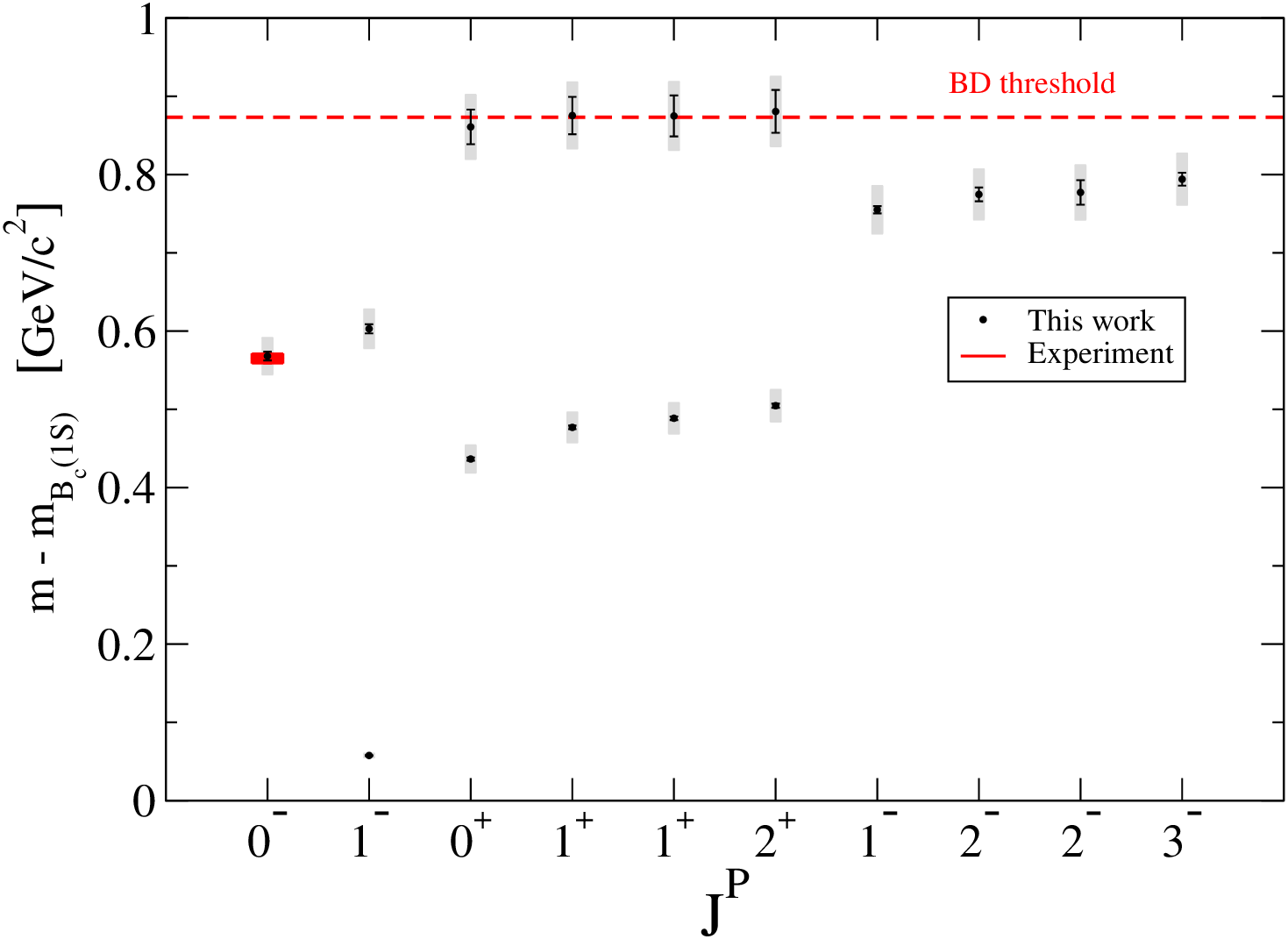}
\caption{Spectrum of $B_c$ meson mass differences with respect to the lightest $B_c$ meson. The red band is an experimental value\cite{Aad:2014laa}. Black points with errors bars are lattice data with statistical errors only. Grey bands are combined statistical and systematic lattice spacing uncertainty, added in quadrature. Numerical values are given in Table~\ref{table-spectrum-Bc}.}
\label{Bc-spectrum}
\end{figure}

The spectrum of $B_c$ mass differences with respect to the lightest $B_c$ state for $1S$, $2S$, $1P$, $2P$ and $1D$ is shown in Fig.~\ref{Bc-spectrum}. The $J^P=0^-$ radially excited S-wave state agrees with a recent observation by the ATLAS Collaboration \cite{Aad:2014laa}. First lattice results are given for the $B_c$ D-wave ground states and the P-wave radial excitations. The $B_c$ results are all under or very near the 
$BD$ 
threshold. Precise $B_c$ spin splittings for $1S$, $2S$, $1P$ and $1D$ are given in Table~\ref{table-Bmeson-compare}. Note that spin splittings are only given between spin 0 and spin 2 for P waves, and spin 1 and spin 3 for D waves, because of the unresolved mixing between the two spin-1 P waves and the two spin-2 D waves.

\begin{table}[tb]
\caption{Comparison of lattice results for $B$, $B_s$ and $B_c$. All quantities are in units of MeV$/c^2$. Only statistical errors are shown for this work. \\ $^a$ Mixing with two-meson scattering states was not calculated. \\ $^b$ There is unresolved mixing between the two $1^+$ states.}
\label{table-Bmeson-compare}
\begin{tabular}{lllllll}
\hline\hline
                            & this work          & HPQCD                      & Lang \textit{et. al.}   & ALPHA     & experiment \\
                            & & \cite{Dowdall:2012ab,Gregory:2010gm} & \cite{Lang:2015hza} & \cite{Bernardoni:2015nqa} & \cite{PDG,Aad:2014laa} \\
\hline
$B_s-B$                     & 84(5)              & 84(2)                      & 81.5(4.1)(1.2) & 88.9(5.7)(2.3)      & 87.35(23)   \\
$B^*-B$                     & 41(3)              & 50(3)                      & 47(7)(1)       & 41.7(4.2)(3.2)(0.3) & 45.0(4)     \\
$B_s^*-B_s$                 & 46.9(5)            & 52(3)                      & 47.1(1.5)(0.7) & 37.8(3.3)(5.8)      & 48.7(2.2)   \\
$B_c^*-B_c$                 & 57.5(3)            & 54(3)                      & -              & -                   & -           \\
$B^*(2S)-B(2S)$             & 60(25)$^a$      & -                          & -              & -                   & -           \\
$B_s^*(2S)-B_s(2S)$         & 48(9)$^a$       & -                          & -              & -                   & -           \\
$B_c^*(2S)-B_c(2S)$         & 35(2)              & 29(26)                     & -              & -                   & -           \\
$B_2^*-B_0^*$               & 148(17)$^a$     & -                          & -              & -                   & -           \\
$B_{s2}^*-B_{s0}^*$         & 104(7)$^a$      & -                          & 142(26)        & -                   & -           \\
$B_{c2}^*-B_{c0}^*$         & 68(3)              & -                          & -              & -                   & -           \\
$B_{c3}^*-B_c^*(1D)$        & 39(10)             & -                          & -              & -                   & -           \\
$B(2S)-B$                   & 617(42)$^a$     & -                          & -              & 791(73)$^a$      & -           \\
$B^*(2S)-B^*$               & 636(39)$^a$     & -                          & -              & -                   & -           \\
$B_s(2S)-B_s$               & 594(14)$^a$     & -                          & -              & 566(57)$^a$      & -           \\
$B_s^*(2S)-B_s^*$           & 595(15)$^a$     & -                          & -              & -                   & -           \\
$B_c(2S)-B_c$               & 568(6)             & 616(19)(1)                 & -              & -                   & 565(4)(5)   \\
$B_c^*(2S)-B_c^*$           & 545(6)             & 591(18)(1)                 & -              & -                   & -           \\
$B_{s0}^*$                  & 5770(6)$^a$     & 5752(16)(5)(25)$^a$     & 5711(13)(19)   & -                   & -           \\
$B_{s1}$                    & 5822(5)$^{a,b}$ & 5806(15)(5)(25)$^{a,b}$ & 5750(17)(19)   & -                   & -           \\
$B_{s1}^\prime$             & 5844(6)$^{a,b}$ & -                          & 5831(9)(6)     & -                   & 5828.7(4)   \\
$B_{s2}^*$                  & 5874(7)$^a$     & -                          & 5853(11)(6)    & -                   & 5839.96(20) \\
$B_{c0}^*-B_c$              & 436(2)             & 429(13)                    & -              & -                   & -           \\
$B_{c1}-B_c^*$              & 419(2)$^b$         & 410(13)$^b$                & -              & -                   & -           \\
$B_{c1}^\prime-B_c^*$       & 431(2)$^b$         & -                          & -              & -                   & -           \\
\hline\hline
\end{tabular}
\end{table}

\begin{figure}[tb]
\centering
\includegraphics[scale=0.6,clip=true]{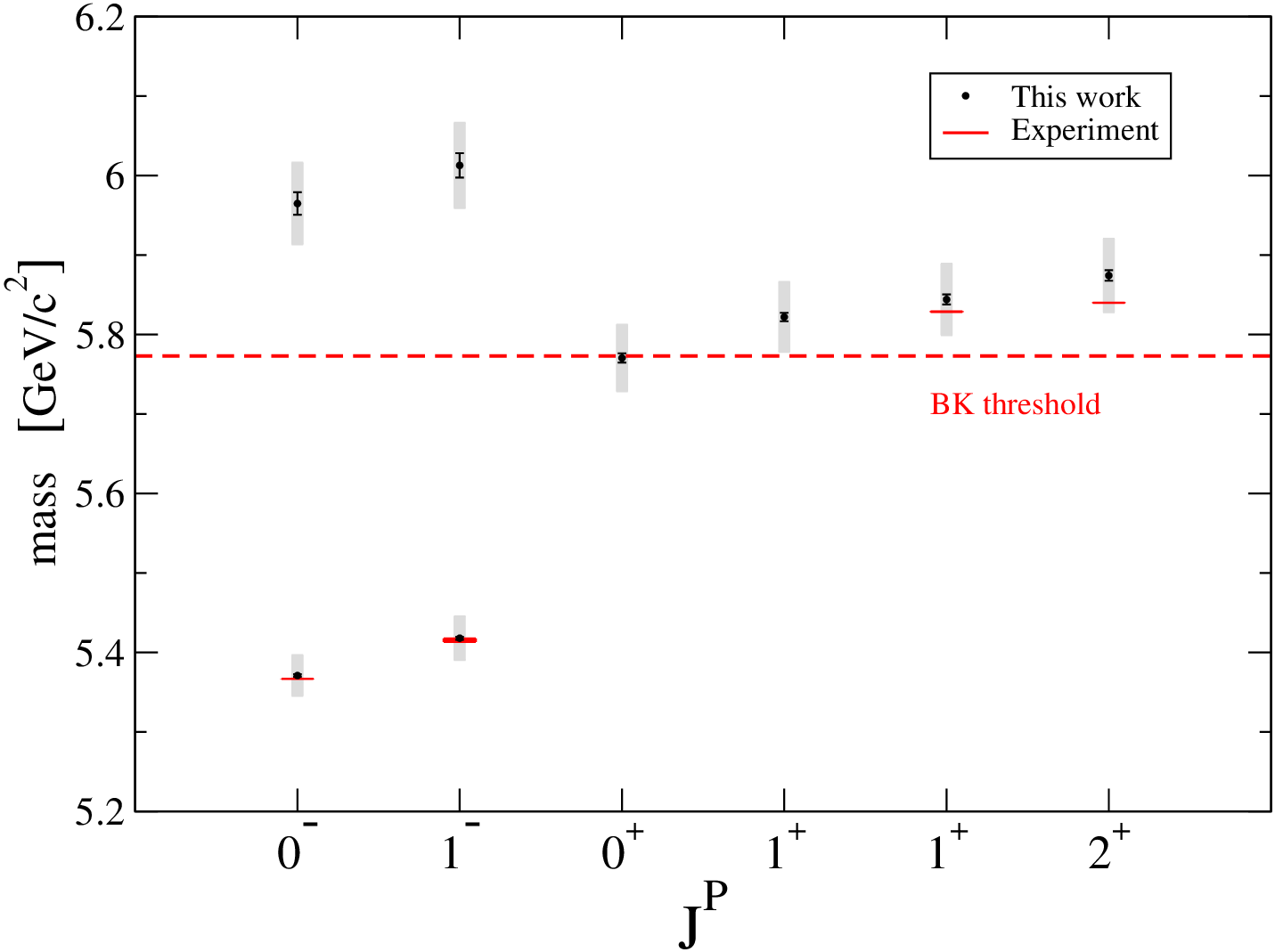}
\caption{Mass spectrum of the $B_s$ meson. Red bands are experimental values. Black points with errors bars are lattice data with statistical errors only. Grey bands are combined statistical and systematic lattice spacing uncertainty, added in quadrature. Numerical values can be found in Table~\ref{table-spectrum-Bs}.}
\label{Bs-spectrum}
\end{figure}

The spectrum of $B_s$ masses for $1S$, $2S$ and $1P$ is shown in Fig.~\ref{Bs-spectrum}.
$B_s$ spin splittings are given in Table~\ref{table-Bmeson-compare}.
The $B_s$ radial and orbital excitations are either very near or above the threshold for breakup into 
$BK$ or $B^*K$. 
Therefore, the $2S$ and $1P$ $B_s$ states have the possibility to mix with these two-meson scattering states. A cross-correlation matrix of $B_s$ with two-meson operators and application of the variational method, which was done in Ref.~\cite{Lang:2015hza}, would be necessary to analyze the mixing and account for its effect.

Since the $J^P=2^+$ state does not decay to 
$BK$
 via S wave, and the S-wave decay to 
$B^*K$
 is suppressed for the $J^P=1^+$ state with $j=\frac{3}{2}$ \cite{Lang:2015hza,Isgur:1991wq}, the two-meson decay must contain nonzero momentum. Therefore the finite-volume thresholds for these two states are higher than the continuum threshold. The same thing applies for the S-wave radial excitations, which are forbidden 
to decay to $B^{(*)}K$
 by S wave. Given the minimum nonzero momentum for the lattice ensemble used in this study, the threshold is around 
5.96 GeV for $B_{s2}^*$ and $B_s^*(2S)$, and 6.01 GeV 
for $B_{s1}^\prime$ and $B_s(2S)$. Recall the systematic uncertainty due to lattice spacing: if the PACS-CS lattice spacing from Eq.~(\ref{aORIG}) is used, then the $B_{s2}^*-B_s$ and $B_{s1}^\prime-B_s$ mass differences agree with Ref.~\cite{Lang:2015hza} and with experiment.

The $0^+$ and $1^+$ states for $B_s$ have not been observed by experiment. Our results agree with Ref.~\cite{Gregory:2010gm} which did not calculate mixing with two-meson states, but differ from Ref.~\cite{Lang:2015hza} which included two-meson states explicitly, as shown in Table~\ref{table-Bmeson-compare}.
This is another indication that including two-meson operators is a necessary step for obtaining reliable masses near and above threshold.

\begin{figure}[tb]
\centering
\includegraphics[scale=0.6,clip=true]{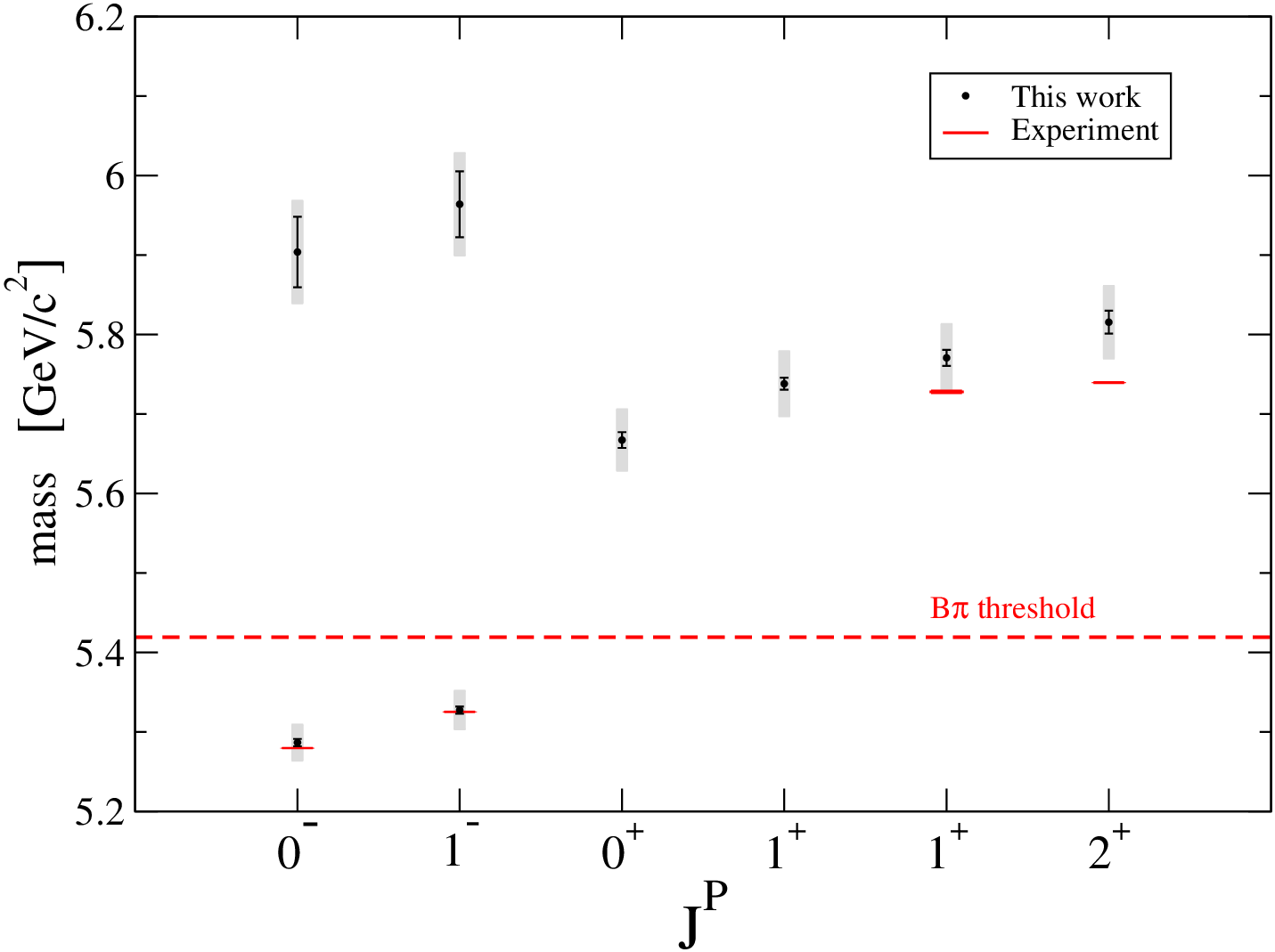}
\caption{Mass spectrum of the $B$ meson. Red bands are experimental values. Black points with errors bars are lattice data with statistical errors only. Grey bands are combined statistical and systematic lattice spacing uncertainty, added in quadrature. Numerical values are given in Table~\ref{table-spectrum-B}.}
\label{B-spectrum}
\end{figure}

The spectrum of $B$ masses for $1S$, $2S$ and $1P$ is shown in Fig.~\ref{B-spectrum}. The LHCb Collaboration has evidence for two states $B(5840)$ and $B(5960)$ in the mass range consistent with $B(2S)$ and $B^*(2S)$ \cite{Aaij:2015qla}. The CDF Collaboration also has evidence for a state $B(5970)$ in this energy range \cite{Aaltonen:2013atp}. $B$ spin splittings are given in Table~\ref{table-Bmeson-compare}. The $B$ radial and orbital excitations are all well above the continuum 
$B\pi$
 threshold. As discussed above for the case of $B_s$, the two-meson S-wave decays are suppressed for $B_2^*$ and $B_1^\prime$ \cite{Isgur:1991wq} and are forbidden for $B^*(2S)$ and $B(2S)$. Given the minimum nonzero lattice momentum used in this study, the threshold for $B_2^*$ and $B^*(2S)$ is 
5.76 GeV, and for $B_1^\prime$ and $B(2S)$ is 5.81 GeV. 
A thorough treatment of the $B$ meson spectrum, like that done in Ref.~\cite{Mohler:2012na} for the $D$ meson P waves, would require the inclusion of two-meson operators.

\section{Conclusions}

In this work, the use of free-form smearing has allowed a lattice study of several bottom mesons and bottomonium states.  The method permits the user to build a source operator with any desired shape, and does not require gauge fixing.  New to the present work is a ``minimal-path'' method for free-form smearing that reduces the computational cost by $O(L)$ on $L^3\times T$ lattices.  Moreover, any number of unrelated smearing choices for a particular field, i.e.\ Eq.~(\ref{freeform}), can be built by reusing a single sum over the shortest link paths, i.e.\ Eq.~(\ref{shortestpaths}).

Beginning with local operators designed to have the appropriate $J^{PC}$ quantum numbers for each hadron, free-form smearing was used to produce a particular hydrogen-like wave-function shape in each case, with the radius and node positions tuned for optimal results.  For each meson, the free-form smearing was applied to the bottom quark at the source, leaving the antiquark ($\bar u/\bar d$, $\bar s$, $\bar c$ or $\bar b$) unsmeared.  Bottom quarks were handled with NRQCD and all other quarks were relativistic.

A nearly complete picture of the bottom meson and bottomonium spectrum below the two-meson breakup threshold is given in Figs.~\ref{bottomonium-spectrum}, \ref{Bc-spectrum}, \ref{Bs-spectrum} and \ref{B-spectrum}
with first lattice results for bottomonium D-wave radial excitations, $B_c$ D-wave ground states and $B_c$ P-wave radial excitations.
The results for states that are well below the breakup threshold (e.g., $BB^{(*)}$ for bottomonium,  $B^{(*)}D$ for $B_c$)  
agree reasonably well with available experimental data and several predictions were made.

Some states that were near or even above threshold are also calculated. These include the $B$ and $B_s$ S-wave radial excitations and P-wave ground states, the $B_c$ P-wave radial excitations and the bottomonium G-wave ground states. Where mixing with meson scattering states may be important, a more complete analysis will be required to obtain robust predictions.
This work, aimed at constructing correlation functions with good radial and orbital excitation signals, is a step in that direction.

The original free-form smearing \cite{vonHippel:2013yfa} was considered unusable for sink smearing. The speedup obtained through our minimal-path implementation gets us closer to the possibility of sink smearing and is an avenue for further development which would make the technique more generally applicable. There are also possibilities to combine this with other ideas like the variational method. A specific application for further development is the calculation of matrix elements, such as in three-point functions where typically one uses a fixed source-sink separation.

\section*{Acknowledgments}

The authors thank Georg von Hippel for helpful discussions about free-form smearing, Martin L\"uscher for making his DD-HMC code publicly available, and the PACS-CS Collaboration for making their dynamical gauge field ensembles publicly available. This work was supported in part by the Natural Sciences and Engineering Research Council (NSERC) of Canada, and by computing resources of WestGrid \cite{westgrid} and SHARCNET \cite{sharcnet}.

\appendix

\section{Additional tables}

Numerical values from Figs.~\ref{bottomonium-spectrum}, \ref{Bc-spectrum}, \ref{Bs-spectrum} and \ref{B-spectrum} are provided in Tables \ref{table-spectrum-bottomonium}, \ref{table-spectrum-Bc}, \ref{table-spectrum-Bs} and \ref{table-spectrum-B} respectively.

\clearpage

\begin{table}[tb]
\caption{Bottomonium simulation energies in lattice units and masses in units of MeV$/c^2$. Lattice values in physical units for the $J=2,3$ states are dimensionally averaged over the different lattice irreducible representations. Only statistical errors are shown for this work.}
\label{table-spectrum-bottomonium}
\begin{tabular}{lllll}
\hline\hline
particle \qquad        & $J^{PC}$ & simulation energy & \multicolumn{2}{c}{mass [MeV$/c^2$]} \\
\cline{4-5}
                       &          &                   & this work   & experiment \cite{PDG,Aaij:2014hla} \\
\hline
$\eta_b(1S)$           & $0^{-+}$ & 0.23401(7)        & 9402.3(3)   & 9398.0(3.2)  \\
$\Upsilon(1S)$         & $1^{--}$ & 0.25963(10)       & 9460.30(26) & 9460.30(26)  \\
$\eta_b(2S)$           & $0^{-+}$ & 0.497(3)          & 9998(6)     & 9999(4)      \\
$\Upsilon(2S)$         & $1^{--}$ & 0.507(3)          & 10020(7)    & 10023.3(3)   \\
$\eta_b(3S)$           & $0^{-+}$ & 0.637(19)         & 10314(44)   & -            \\
$\Upsilon(3S)$         & $1^{--}$ & 0.646(23)         & 10334(52)   & 10355.2(5)   \\
$h_b(1P)$              & $1^{+-}$ & 0.4540(5)         & 9900.2(9)   & 9899.3(1.0)  \\
$\chi_{b0}(1P)$        & $0^{++}$ & 0.4386(4)         & 9865.3(1.0) & 9859.4(5)    \\
$\chi_{b1}(1P)$        & $1^{++}$ & 0.4507(5)         & 9892.7(9)   & 9892.8(4)    \\
$\chi_{b2}(1P)$        & $2^{++}$ & 0.4599(6)$_E$     & 9913.1(1.0) & 9912.2(4)    \\
                       &          & 0.4596(6)$_{T_2}$ &             &              \\
$h_b(2P)$              & $1^{+-}$ & 0.595(5)          & 10219(11)   & 10259.8(1.2) \\
$\chi_{b0}(2P)$        & $0^{++}$ & 0.584(5)          & 10194(12)   & 10232.5(6)   \\
$\chi_{b1}(2P)$        & $1^{++}$ & 0.592(5)          & 10212(11)   & 10255.5(5)   \\
$\chi_{b2}(2P)$        & $2^{++}$ & 0.600(6)$_E$      & 10227(13)   & 10268.7(5)   \\
                       &          & 0.598(6)$_{T_2}$  &             &              \\
$\eta_{b2}(1D)$        & $2^{-+}$ & 0.571(3)$_E$      & 10163(3)    & -            \\
                       &          & 0.5693(15)$_{T_2}$&             &              \\
$\Upsilon  (1D)$       & $1^{--}$ & 0.5646(9)         & 10150(2)    & -            \\
$\Upsilon_2(1D)$       & $2^{--}$ & 0.5689(15)        & 10161(3)    & 10163.7(1.4) \\
                       &          & 0.5697(17)        &             &              \\
$\Upsilon_3(1D)$       & $3^{--}$ & 0.5728(18)$_{A_2}$& 10172(3)    & -            \\
                       &          & 0.5761(10)$_{T_1}$&             &              \\
                       &          & 0.5730(16)$_{T_2}$&             &              \\
$\eta_{b2}(2D)$        & $2^{-+}$ & 0.712(11)$_E$     & 10458(26)   & -            \\
                       &          & 0.693(18)$_{T_2}$ &             &              \\
$\Upsilon  (2D)$       & $1^{--}$ & 0.675(17)         & 10401(39)   & -            \\
$\Upsilon_2(2D)$       & $2^{--}$ & 0.690(16)$_E$     & 10447(25)   & -            \\
                       &          & 0.699(10)$_{T_2}$ &             &              \\
$\Upsilon_3(2D)$       & $3^{--}$ & 0.703(24)$_{A_2}$ & 10459(37)   & -            \\
                       &          & 0.704(22)$_{T_1}$ &             &              \\
                       &          & 0.698(15)$_{T_2}$ &             &              \\
$h_{b3}(1F)$           & $3^{+-}$ & 0.654(4)$_{A_2}$  & 10355(7)    & -            \\
                       &          & 0.655(4)$_{T_2}$  &             &              \\
$\chi_{b3}(1F)$        & $3^{++}$ & 0.653(5)          & 10350(12)   & -            \\
$\eta_{b4}(1G)$        & $4^{-+}$ & 0.749(4)          & 10568(9)    & -            \\
$\Upsilon_4(1G)$       & $4^{--}$ & 0.760(9)          & 10592(21)   & -            \\
\hline\hline
\end{tabular}
\end{table}

\clearpage

\begin{table}[tb]
\caption{$B_c$ meson simulation energies in lattice units and mass differences with respect to the lightest $B_c$ state in units of MeV$/c^2$. Lattice values in physical units for the $J=2,3$ states are dimensionally averaged over the different lattice irreducible representations. Only statistical errors are shown for this work.}
\label{table-spectrum-Bc}
\begin{tabular}{lllll}
\hline\hline
particle            & $J^P$ & simulation energy  & \multicolumn{2}{c}{$m-m_{B_c(1S)}$ [MeV$/c^2$]} \\
\cline{4-5}
                    &       &                    & this work & experiment \cite{PDG,Aad:2014laa} \\
\hline
$B_c(1S)$           & $0^-$ & 0.73373(14)        & 0         & 0 \\
$B_c^*(1S)$         & $1^-$ & 0.75914(14)        & 57.5(3)   & - \\
$B_c(2S)$           & $0^-$ & 0.985(2)           & 568(6)    & 565(6) \\
$B_c^*(2S)$         & $1^-$ & 1.000(3)           & 603(6)    & - \\
$B_{c0}^*(1P)$      & $0^+$ & 0.9266(8)          & 436(2)    & - \\
$B_{c1}(1P)$        & $1^+$ & 0.9445(9)          & 477(2)    & - \\
$B_{c1}^\prime(1P)$ & $1^+$ & 0.9496(10)         & 489(2)    & - \\
$B_{c2}^*(1P)$      & $2^+$ & 0.9564(13)$_E$     & 505(3)    & - \\
                    &       & 0.9569(13)$_{T_2}$ &           &   \\
$B_{c0}^*(2P)$      & $0^+$ & 1.114(9)           & 861(22)   & - \\
$B_{c1}(2P)$        & $1^+$ & 1.121(10)          & 875(24)   & - \\
$B_{c1}^\prime(2P)$ & $1^+$ & 1.120(11)          & 875(26)   & - \\
$B_{c2}^*(2P)$      & $2^+$ & 1.127(12)$_E$      & 881(27)   & - \\
                    &       & 1.120(13)$_{T_2}$  &           &   \\
$B_c^*(1D)$         & $1^-$ & 1.0674(18)         & 755(5)    & - \\
$B_{c2}(1D)$        & $2^-$ & 1.074(3)$_E$       & 775(9)    & - \\
                    &       & 1.077(5)$_{T_2}$   &           &   \\
$B_{c2}^\prime(1D)$ & $2^-$ & 1.072(14)$_E$      & 777(16)   & - \\
                    &       & 1.080(7)$_{T_2}$   &           &   \\
$B_{c3}^*(1D)$      & $3^-$ & 1.089(8)$_{A_2}$   & 794(8)    & - \\
                    &       & 1.085(2)$_{T_1}$   &           &   \\
                    &       & 1.083(6)$_{T_2}$   &           &   \\
\hline\hline
\end{tabular}
\end{table}

\clearpage

\begin{table}[tb]
\caption{$B_s$ meson simulation energies in lattice units and masses in units of MeV$/c^2$. Lattice values in physical units for the $J=2$ states are dimensionally averaged over the different lattice irreducible representations. Only statistical errors are shown for this work.}
\label{table-spectrum-Bs}
\begin{tabular}{lllll}
\hline\hline
particle            & $J^P$ & simulation energy & \multicolumn{2}{c}{mass [MeV$/c^2$]} \\
\cline{4-5}
                    &       &                   & this work   & experiment \cite{PDG} \\
\hline
$B_s(1S)$           & $0^-$ & 0.4130(3)         & 5370.9(1.6) & 5366.77(24) \\
$B_s^*(1S)$         & $1^-$ & 0.4337(4)         & 5417.8(1.8) & 5415.4(2.3) \\
$B_s(2S)$           & $0^-$ & 0.675(6)          & 5965(14)    & -           \\
$B_s^*(2S)$         & $1^-$ & 0.697(7)          & 6013(15)    & -           \\
$B_{s0}^*(1P)$      & $0^+$ & 0.590(2)          & 5770(6)     & -           \\
$B_{s1}(1P)$        & $1^+$ & 0.612(2)          & 5822(5)     & -           \\
$B_{s1}^\prime(1P)$ & $1^+$ & 0.622(3)          & 5844(6)     & 5828.7(4)   \\
$B_{s2}^*(1P)$      & $2^+$ & 0.635(3)$_E$      & 5874(7)     & 5840.0(2)   \\
                    &       & 0.636(3)$_{T_2}$  &             &             \\
\hline\hline
\end{tabular}
\end{table}

\begin{table}[tb]
\caption{$B$ meson simulation energies in lattice units and masses in units of MeV$/c^2$. Lattice values in physical units for the $J=2$ states are dimensionally averaged over the different lattice irreducible representations. Only statistical errors are shown for this work.}
\label{table-spectrum-B}
\begin{tabular}{lllll}
\hline\hline
particle         & $J^P$ & simulation energy & \multicolumn{2}{c}{mass [MeV$/c^2$]} \\
\cline{4-5}
                 &       &                   & this work & experiment \cite{PDG,Aaltonen:2013atp,Aaij:2015qla} \\
\hline
$B(1S)$          & $0^-$ & 0.3757(20)        & 5287(5)   & 5276.26(17) \\
$B^*(1S)$        & $1^-$ & 0.3937(19)        & 5327(4)   & 5325.2(4)   \\
$B(2S)$          & $0^-$ & 0.648(20)         & 5904(44)  & -           \\
$B^*(2S)$        & $1^-$ & 0.675(18)         & 5964(41)  & -           \\
$B_0^*(1P)$      & $0^+$ & 0.544(4)          & 5667(10)  & -           \\
$B_1(1P)$        & $1^+$ & 0.575(3)          & 5738(8)   & -           \\
$B_1^\prime(1P)$ & $1^+$ & 0.590(4)          & 5771(10)  & 5727.7(1.6) \\
$B_2^*(1P)$      & $2^+$ & 0.609(7)$_E$      & 5815(14)  & 5739.4(5)   \\
                 &       & 0.610(6)$_{T_2}$  &           &             \\
\hline\hline
\end{tabular}
\end{table}

\end{document}